\documentclass[letterpaper,dvipsnames]{article}
\pdfoutput=1 

\usepackage{jheppub} 
\usepackage{hyperref}

\usepackage{amsmath}
\usepackage[T1]{fontenc} 
\usepackage{braket}
\usepackage{subcaption}
\usepackage{array}
\usepackage{xcolor}
\usepackage{slashed}
\usepackage{comment}
\usepackage[many]{tcolorbox}    	

\newtcolorbox{boxA}{
    fontupper = \bf,
    boxrule = 1.5pt,
    colframe = black 
}

\newtcolorbox{boxE}{
    enhanced, 
    boxrule = 0pt, 
    borderline = {0.75pt}{0pt}{main}, 
    borderline = {0.75pt}{2pt}{sub} 
}

\newcommand{\PQ}{U(1)^{(0)}_{\sf PQ}}
\newcommand{\PQB}{U(1)^{(2)}}
\newcommand{\W}{U(1)^{(1)}_e}
\newcommand{\tH}{U(1)^{(1)}_m}
\newcommand{\Lam}{\mathcal{L}_{\rm axMax}}

\usepackage[compat=1.1.0]{tikz-feynman}

\title{The Monodromic Axion-Photon Coupling}
\author{Prateek Agrawal}
\author{and Arthur Platschorre}

\affiliation{Rudolf Peierls Centre for Theoretical Physics, University of Oxford, Parks Road, Oxford OX1 3PU, United Kingdom}

\abstract{
We consider the general form of the axion coupling to photons in the axion-Maxwell theory. On general grounds this coupling takes the form of a monodromic function of the axion, which we call $g(a)$, multiplying the Chern-Pontryagin density $F \widetilde{F}$ of the photon. We show that the non-linearity of $g(a)$ is a spurion for the shift symmetry of the axion. In this context, when $g(a) \neq \mathbb{Z}a$, the linearized coupling of the axion $g'(a)$ is not quantized and there is a correlated mass term for the axion. Singularities in $g(a)$ due to the fast rearrangement of degrees of freedom are shown to have corresponding cusps and singularities in the axion potential.
We derive the general form of $g(a)$ for the QCD axion, axions with perturbatively broken shift symmetries and axions descending from extra dimensions. In all cases, we show that there is a uniform general form of the monodromic function $g(a)$ and it is connected to the axion potential. 
}

\begin{document} 

\maketitle


\section{Introduction}
The axion-Maxwell Lagrangian describes the low-energy physics of one of the most compelling new physics candidates, the axion, and its experimentally important coupling to photons. The discovery of the axion-photon interaction will not just be a discovery of a new particle, but can provide deep insights into the structure of the standard model. The QCD axion elegantly explains the non-observation of CP violation in the strong sector~\cite{Peccei:1977hh,Weinberg:1977ma,Wilczek:1977pj}. Axions can solve the cosmological puzzle of dark matter~\cite{Preskill:1982cy,Abbott:1982af,Dine:1982ah} and may appear as dark energy \cite{Marsh:2012nm,PhysRevD.83.123526}. The axion-photon coupling can provide access to the fundamental unit of electric charge~\cite{Agrawal:2017cmd,Fraser:2019ojt,Agrawal:2019lkr,Agrawal:2020euj} and test simple models of Grand Unification~\cite{Agrawal:2022lsp}. Axions have a strong interplay with ideas in quantum gravity~\cite{Heidenreich:2021yda,Reece:2023czb} and string theory~\cite{Svrcek:2006yi,Arvanitaki:2009fg,Cicoli:2012sz}.

A large part of this wealth of information derives from the special nature of the axion-photon coupling and the associated symmetries and redundancies. In this work, we derive the general form of this coupling that is ideally suited to study the quantization of the axion-photon coupling,  the physics of axion domain walls and strings and the symmetry structure in axion-Maxwell theory. 

We argue that the general low-energy axion-Maxwell Lagrangian takes the form,
\begin{align}
    \Lam
    &=
    -\frac{1}{4e^2} F_{\mu\nu} F^{\mu\nu}
    +\frac12 F_a^2 \partial_\mu a \partial^\mu a - V(a)
    +
    \frac{g(a)}{16\pi^2} 
     F_{\mu\nu}
     \widetilde{F}^{\mu\nu} 
     \,.
\end{align}
For convenience we have chosen a basis where $F_a$, the fundamental period of the axion, as well as the electromagnetic gauge coupling $e$ are included in the kinetic term. The function $g(a)$ is a monodromic function, defined by the property,
\begin{align}
    g(a + 2\pi)
    &=
    g(a) + 2\pi n  \,.
\end{align}
The integer $n$ is the monodromic charge of the function $g(a)$. This property of the monodromic function arises from the discrete gauge symmetry of the axion, $a \to a + 2\pi$, under which the path integral weight, $e^{iS}$, is required to be invariant. 

It is an extremely important fact that the monodromy of $g(a)$ does not imply that the perturbative coupling of the axion to photons around the {\sf CP} conserving point $a=0$ is quantized. Indeed, the coupling for canonically normalized fields,
\begin{align}
    g_{a\gamma\gamma}
    &=
    \frac{\alpha_{\rm em}}{\pi F_a} g'(a)|_{a=0},
    \label{quantizationiscausation}
\end{align}
which can be an arbitrary number for a non-linear monodromic function $g(a)$. 

This resolves a small puzzle in the QCD axion coupling to photons, as was noted in \cite{Agrawal:2017cmd} and further discussed in \cite{Fraser:2019ojt}. On one hand, we usually justify the non-quantized couplings of the axion by invoking the mixing with the pion. On the other hand, for all values of the axion the pion remains heavy and can stay integrated out, leaving an apparent non-monodromic function. The resolution to the non-quantization of the coupling therefore should appear in the low-energy axion-Maxwell theory without needing to invoke the pion. Indeed, this is achieved by a monodromic non-linear function $g(a)$.

The form of the coupling $g(a)$ generated by the anomaly between $U(1)_{\sf PQ}$ and $U(1)_{\rm em}^2$ is $g(a) = n a$. This form of $g(a)$ is protected by the continuous shift symmetry of the axion, which also protects the axion from getting a mass. Both the potential $V(a)$ and a non-linear $g(a)$ are therefore spurions for the axion continuous shift symmetry breaking \cite{Agrawal:2017cmd,Fraser:2019ojt}. This gives a precise sense in which the deviation from quantization of axion couplings and the generation of a mass are linked.  Thus, while the monodromy of $g(a)$ follows from topology, the special case of $g(a)= n a$ additionaly requires the presence of a continuous global shift symmetry.

In general we expect the size of the two spurions for the same symmetry to be commensurate. If the axion-photon coupling is nearly quantized, then we can express the degree of non-quantization as
\begin{align}
g(a)-n a = z f(a),
\label{eq:nonqg}
\end{align}
with $z \ll 1$ and $f(a)$ an $O(1)$ periodic function. The estimate for the mass of the axion is
\begin{align}
    m^{2}
    &\sim
    z\frac{\Lambda^4}{F_{a}^{2}} 
    \,.
\end{align}
where $\Lambda$ is the UV cutoff of the effective theory consistent with the coupling $g(a)$ (e.g.~for the QCD axion $\Lambda \simeq \Lambda_{\rm QCD}$).
We emphasize that this is a heuristic estimate and the actual correlation may be different in specific examples. However this correlation highlights the point that if there is an axion that is parametrically lighter than its naively expected mass, that also corresponds to a coupling to photons that is very nearly an integer. Similarly, if an axion coupled to photons picks up a mass it generically also picks up a non-linear $g(a)$ coupling to photons \cite{Fraser:2019ojt}. 

The general periodic function $f(a)$ in equation~\eqref{eq:nonqg} can be expanded in Fourier modes. In some cases only the first few terms in the expansion dominate. This is simply the expected contribution from axion-dependent perturbative corrections to non-topological quantities like $\alpha_{\rm em}$. However, in many cases, including the case of the QCD axion, the final form of $g(a)$ requires the sum over the entire Fourier tower, and it is interesting that a closed form for $g(a)$ can be derived. 
 
The functional coupling $g(a)$ elucidates many interesting physics points. As mentioned above, it captures the correct monodromy in the axion-Maxwell Lagragian when all other fields can be integrated out for all values of the axion. In cases where this is not possible (e.g.~when some particles become light at some value of the axion field) $g(a)$ also captures fast rearrangement of degrees of freedom through its singularities at isolated points. This correlates with cusps and singularities in the axion potential at the same point, and interesting dynamics induced on an axion domain wall.

Phenomenologically, the full non-linear form of $g(a)$ is most relevant for scenarios where the axion traverses an $O(1)$ fraction of its field range. This is certainly true for axion strings and domain walls, and  sharp features in $g(a)$ can affect axion emission from these objects. It can also be true for dense axion objects, like axion miniclusters or superradiant axion clouds surrounding rotating black holes.

The fact that in many simple models the whole Fourier tower needs to be summed up to get the relevant $g(a)$ highlights another interesting point. For effective field theories involving compact fields the standard polynomial basis might not be the most convenient basis to work in.

This paper is organised as follows. In section \ref{sec3:generalproperties}, the general properties  of $g(a)$ and the symmetries of the axion-Maxwell Lagrangian in the presence of $g(a)$ are discussed, together with the connection between the mass and non-quantization of $g(a)$. Section \ref{sec:QCDaxion}
discusses the QCD axion and the corresponding axion-photon coupling. In section \ref{PQbreaking}, the important case of case of perturbative shift symmetry breaking is introduced and shown to share many features of the QCD axion. The final section \ref{sec:5dfermions} is entirely devoted to axion potentials and photon couplings in the presence of a tower of states.

\section{Axion-Maxwell Theory}\label{sec3:generalproperties}
In this section, we discuss the general properties and symmetries of the axion-Maxwell Lagrangian $\Lam$ in the presence of an effective axion-photon coupling $g(a)$.
\begin{align}
    \Lam
    &=
    -\frac{1}{4e^2} F_{\mu\nu} F^{\mu\nu}
    +\frac12 F_a^2 \partial_\mu a \partial^\mu a - V(a)
    +
    \frac{g(a)}{16\pi^2} 
     F_{\mu\nu}
     \widetilde{F}^{\mu\nu}  \,.
     \label{lowenergyeft}
\end{align}
Here $F_{a}$ is the fundamental period of the axion, and we have normalized the gauge field such that the electric charge of the electron is -1. 

\subsection{Quantization}
The function $g(a)$ is a monodromic function, defined by the property,
\begin{align}
    g(a + 2\pi)
    &=
    g(a) + 2\pi n  \,.
    \label{eq:quant}
\end{align}
Here $n$ is the monodromic charge of the function $g(a)$, which is usually taken to be an integer in order for the 
path integral weight $e^{iS}$ to be invariant under the identification $a \equiv a + 2\pi$. 

To be more precise, the quantization of the monodromy depends on the global structure of the gauge group \cite{Aharony:2013hda,Tong:2017oea}. If the smallest allowed representation has physical electric charge $eq$, then the electromagnetic instanton number $I = \frac{1}{16 \pi^{2}} \int F \widetilde{F}$ is valued in $\frac{\mathbb{Z}}{q^{2}}$. In such a model, the monodromic charge $n$ can take values in $q^{2} \mathbb{Z}$. Correspondingly, colourless magnetic monopoles can have a physical minimum magnetic charge $q_{m} = \frac{2\pi}{eq}$ by Dirac quantization.

The quantization of the monodromic charge $n$ can also be shown by several connected topological arguments similar to Dirac's argument for quantization of electric charge in $U(1)$ gauge theory. In a theory with a $U(1)$ gauge field and an axion discrete gauge shift symmetry, both magnetic monopoles and axion strings exist as twisted sectors. In quantum field theory the cores of these objects may be singular, but in presence of gravity these singularities will be behind a horizon.

Consider a  magnetic monopole with minimum magnetic charge $q_{m} = \frac{2\pi}{eq}$ scattering on a trajectory through an axion string loop~\cite{Sikivie:1984yz}. For the purposes of this thought experiment, it does not matter if the axion has a mass or not. Along the trajectory, the monopole sees a monodromy of the axion as $g(\theta+2\pi)-g(\theta)$. Through the Witten effect, this implies that the monopole electric charge shifts by $\Delta q_{e} =  -\frac{q_{m} e^{2}}{4 \pi^{2}}  (g(\theta+2\pi)-g(\theta)) = -  \frac{n e}{q}$. By Dirac-Zwanziger quantization of dyons $\Delta q_{e}  \in eq\mathbb{Z}$ and therefore the monodromic charge $n \in q^{2}\mathbb{Z}$.  

There is another argument for the quantization in this example that does not directly rely on the Witten effect. As the monopole traverses the axion string loop, electric charge is carried from the magnetic monopole to the axion string by the Goldstone-Wilczek current~\cite{Goldstone:1981kk}, which is a purely bulk effect following from the axion-photon coupling. The current can be integrated to calculate the total charge exchanged between the monopole and the axion string (see e.g.~\cite{Agrawal:2020euj}). The charge carriers on the axion strings are zero modes of bulk fields (e.g.~PQ fermions), and therefore are also quantized.

 In the remainder of this paper we shall focus on theories in which the smallest unit of charge is that of the electron  such that $g(a)$ has integer monodromy, but we shall briefly return to this issue when we discuss the QCD axion and the allowed standard model representations. Results for other minimal charges can be recovered by the appropriate multiplication.

\subsection{Symmetries of $\Lam$}
We begin by reviewing the symmetry structure of this theory (see e.g.~\cite{Choi:2022fgx,Brennan:2020ehu,Hidaka:2020iaz,Cordova:2022ieu} for further details) in the limit of a massless axion and $g(a) = a$.

In fact, in the even simpler limit where the axion coupling to photon is turned off, we have the following symmetry structure. The Maxwell theory is well-known to have two global one-form symmetries, the electric $\W$ and magnetic $\tH$, under which Wilson lines and `t Hooft lines transform respectively \cite{Gaiotto:2014kfa}. 
These symmetries act on the photon and the dual photon by a shift by a closed one-form,
\begin{align}
A &\xrightarrow{\W} A + c^{(1)} , \quad dc^{(1)} = 0  \,,
\\
\tilde{A} & \xrightarrow{\tH} \tilde{A} + \tilde{c}^{(1)}  , \quad d\tilde{c}^{(1)} = 0  \,.
\end{align}
Equivalently, non-contractible Wilson and `t Hooft loops transform by a $U(1)$ phase under the respective symmetries. In the absence of charged matter, these symmetries above are clearly symmetries of the Maxwell Lagrangian.
In the real world we know the electric symmetry to be emergent below the electron mass, and it is strongly believed that the magnetic symmetry will also be broken completely \cite{Heidenreich:2021xpr}. 

The massless axion Lagrangian has an ordinary global $\PQ$ symmetry, the usual continuous shift symmetry of the massless axion, 
\begin{align}
a \xrightarrow{\PQ} a + c^{(0)} ,\quad dc^{(0)} =0  \,,
\end{align}
as well as a two-form symmetry $\PQB$ which measures the axion winding number, under which axion string worldsheets are charged. This symmetry is a shift symmetry of the dual two-form field $B$,
\begin{align}
B \xrightarrow{\PQB} B + c^{(2)}  , \quad dc^{(2)} = 0  \,.
\end{align}
or equivalently a phase rotation of a non-contractible axion string worldsheet. The symmetry structure at this level is thus,
\begin{align}
    \PQ \times \PQB \times \W \times \tH  \,.
\end{align}

A linear axion-photon coupling in the Lagrangian introduces mixed anomalies between the $\PQ$ and the one-form symmetries of Maxwell theory, as well as an ABJ anomaly~\cite{Adler:1969gk,Bell:1969ts} for the axion shift symmetry,
\begin{align}
\partial_\mu j^\mu_{\sf PQ}
&=
\frac{1}{16 \pi^{2}} 
F_{\mu\nu}
\widetilde{F}^{\mu\nu}  \,.
\label{eq:conservationpq}
\end{align}
Therefore, from this point of view it is somewhat mysterious which symmetry is formally responsible for protecting the axion from getting a mass. One argument could be that if we are working on $\mathbb{R}^4$, then there are no Abelian instantons and the RHS does not produce any physical effect. 
In particular, we can define the PQ charge on a fixed time slice, 
\begin{align}
{\sf Q}
&=
\int d^3 x
\left(
j^0_{\sf PQ}
-\frac{1}{8 \pi^2}
\epsilon^{ijk}A_i \partial_j A_k
\right)  \,.
\label{sec3:conservedcharge}
\end{align}
The charge $\sf{Q}$ defined on $\mathbb{R}^3$ is gauge invariant, so it looks like we can rescue the shift symmetry of the axion if we are content to work on $\mathbb{R}^4$.

However, this argument is a bit too quick. We cannot use the same argument for potential UV contributions to the axion mass. The topology of spacetime seen by the Abelian gauge field can change in the UV, both in extra-dimensional theories and 4D theories, a simple example being the `t Hooft-Polyakov monopole. It will be much more useful to find a symmetry and associated spurions that parametrize both UV and IR mass generation effects on general manifolds. This is especially valuable for the case of axions where we expect at least quantum gravitational effects to generate a mass.

On a general manifold there does not exist a gauge-invariant charge $\sf{Q}$. This is most easily seen if our spatial slice is $S^{1} \times S^{2}$ with magnetic flux $m = \frac{1}{2\pi}\int_{S^{2}} F_{23}$ on the $S^{2}$. Performing a large gauge transformation $A_{1} \rightarrow A_{1} + 2 \pi$ on the compact $S^{1}$ shifts the charge $Q$ by $m$. The operator implementing the $U(1)_{\sf Q}$ symmetry, $\exp(i \alpha {\sf Q})$ with $\alpha \in [0,2\pi)$ is not gauge-invariant under this transformation. A more modern viewpoint is that the introduction of topologically non-trivial backgrounds can be captured by turning $\PQ$ into an unbroken non-invertible symmetry \cite{Choi:2022fgx,Cordova:2022ieu,Shao:2023gho}. 
In the cases that such background fluxes or instantons become dynamical, the symmetry is explicitly broken and the axion is expected to get a mass. This is the case when the axion in addition to the photon is also coupled to a non-Abelian gauge theory or when magnetic monopoles are dynamical.

In the presence of a general effective axion-photon coupling $g(a)$, the conservation equation \eqref{eq:conservationpq} of $\PQ$ is modified to
\begin{equation}
\partial_{\mu} j^{\mu}_{\sf{PQ}} = \frac{g'(a)}{16 \pi^{2}} F_{\mu \nu} \widetilde{F}^{\mu \nu}
\end{equation}
and no such conserved PQ charge (equation \eqref{sec3:conservedcharge}) exists unless $g'(a)$ is an integer. General forms of $g'(a)$ therefore explicitly break $\PQ$. It is this sense in which $g(a)$ can parametrize both the UV dynamical topology changes as well as other dynamical sources of $\PQ$ breaking. The non-linearity of $g(a)$ therefore acts as a spurion for the $\PQ$ shift symmetry.

\subsection{General properties of $g(a)$}

We have seen that the general non-linear function $g(a)$  breaks the (non-invertible) axion shift symmetry. Therefore, we expect a general connection between $g(a)$ and the potential for the axion $V(a)$. Indeed, $g'(a)\notin \mathbb{Z}$ implies a mass for the axion. Similarly, a potential for the axion $V(a)$ and a quantized axion-photon coupling will flow to a non-quantized $g(a)$ with the same monodromy. 

In the examples considered in this paper, the connection between the potential $V(a)$ and axion-photon coupling $g(a)$ is best provided by the repackaging of a real parameter $z$ together with the axion $a$ into a complex quantity,
\begin{equation}
 \mathcal{Z} = z e^{ia} \,.
\end{equation}
The real and imaginary parts of powers of $\mathcal{Z}$ respectively contribute to the CP even potential $V(a)$ and CP odd effective axion-photon coupling $g(a)$, providing the connection between the two. The duality $\mathcal{Z} \rightarrow \frac{1}{\mathcal{Z}}$ leaves both the potential and $g(a)$ invariant. Such a repackaging of the parameters in the case of instanton contributions to the axion potential $z = e^{- S_{\mathrm{inst}}}$ was already noted in contributions to the superpotential in \cite{Dine:1987bq}. 

Common to these examples is a prototypical axion-photon coupling $g(a)$ that can be expressed as a contour integral,
\begin{equation}
g(a) = \mathrm{Im}\int_{C} \frac{d\mathcal{Z}}{\mathcal{Z}} \frac{1-\mathcal{Z}}{1+\mathcal{Z}}
\,,
\label{complexcontour}
\end{equation}
where the contour $C$ is an arc at radius $z$ of angular size $a$. The monodromic charge $n$ can be extracted from equation \eqref{complexcontour} by the poles of the integrand that are included in the closed contour at radius $z$. The poles for this particular function are located at $\mathcal{Z} = 0$ and $\mathcal{Z}= -1$ with respective residues $1$ and $-2$, giving a monodromic charge that is $n = \mathrm{sign}(1-z)$. 

\begin{figure*}[t]
	\centering
	\includegraphics[width=0.6\textwidth]{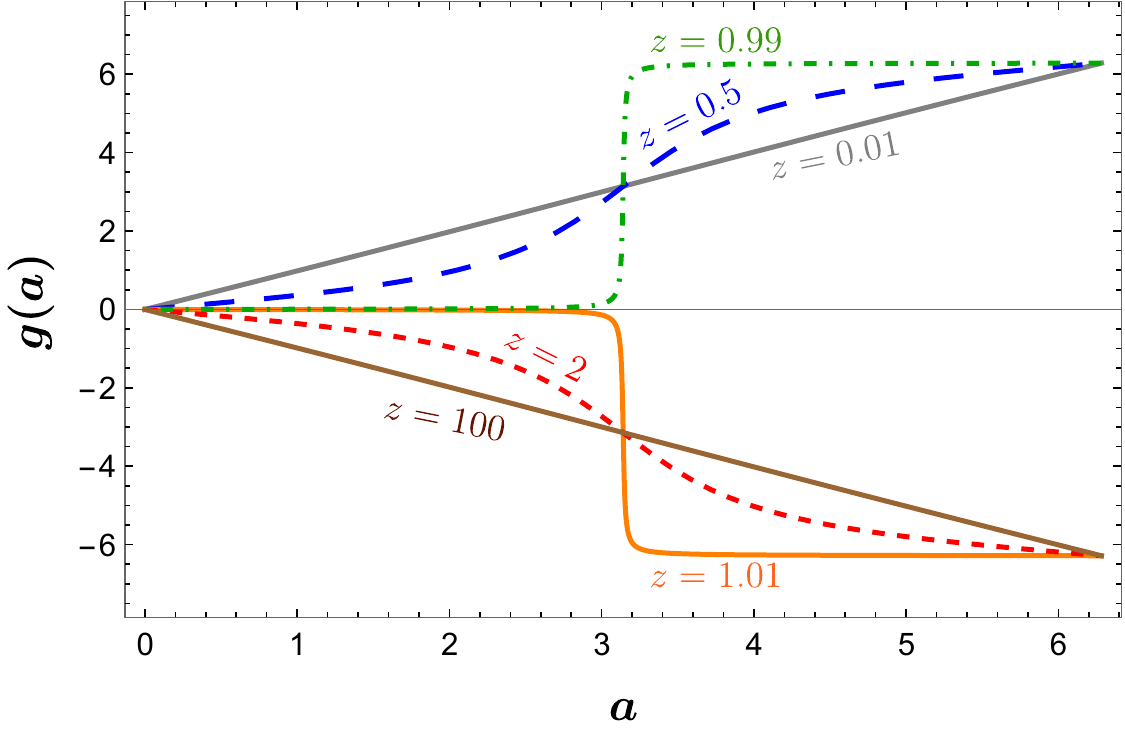}
		\caption{The effective axion-photon coupling $g(a)$ for the prototypical example in equation~\eqref{eq:protoga} at values $z = \{0.01, 0.5, 0.99,1.01, 2, 100\}$ showing that $g(a)$ jumps across $z=1$ and further becomes discontinuous as $z\to1^\pm$ at $a= \pi$.} 
	\label{fig:gaplot}
\end{figure*}

The effective axion-photon coupling  can be extracted from equation \eqref{complexcontour} by performing the contour integral over the arc $C$,
\begin{equation}
g(a) = 2 \arctan{\left(\frac{1-z}{1+z} \tan{\frac{a}{2}} \right)} + 2\pi \mathrm{sign}(1-z)\Theta(a - \pi)  \,,
\label{eq:protoga}
\end{equation}
where $\Theta$ is the Heaviside function. The full profile of $g(a)$ is plotted for several relevant parameter values in figure \ref{fig:gaplot}.
The function $g(a)$ can be decomposed into a monodromic part $na$ and a periodic part, the latter captures the explicit breaking of the continuous axion shift symmetry.



The feature most relevant to  current experiments is the slope of the effective axion-photon coupling around the minimum $a=0$ of the potential,
\begin{equation}
g'(0) = \frac{1-z}{1+z}  \,.
\end{equation}
Under the transformation $z \rightarrow \frac{1}{z}$, the slope and monodromy of $g(a)$ swap signs, which is a reflection of the $\mathcal{Z}\to 1/{\mathcal{Z}}$ duality mentioned above.

There are three values of the real parameter $z$ that are interesting. At the points $z = \{0,\infty\}$, 
$g'(a) \in \mathbb{Z}$ and the axion shift symmetry is restored. In our examples, the axion potential also vanishes for these values of $z$.
The function $g(a)$ does not have a well-defined limit as $z \to 1$, it changes discontinuously across $z=1$. In this limit, $g'(0) = k \in \mathbb{Z}$, but the axion shift symmetry is not restored, and the monodromy is not equal to $k$. 

Common to our examples will be the restoration of a $\mathbb{Z}_{2}$ discrete symmetry at $z=1$, which has an anomaly with electromagnetism. The anomaly is captured in the low-energy effective theory by $g(a)$ changing discontinuously across $z=1$.  
Furthermore, the different profiles  $\lim\limits_{\ {z\to 1^-}}g(a)$ and $\lim\limits_{\ z\to1^+} g(a)$ are both discontinuous at the point $a = \pi$, describing a fast rearrangement of degrees of freedom and restoration of a $U(1)$ symmetry.
This discontinuity in $g(a)$ at $a=\pi$ is reproduced at the same point by a singularity in the potential $V(a)$ or its derivatives.

\section{The QCD axion}\label{sec:QCDaxion}
We study QCD in the two flavour approximation $N_{f} =2$ coupled to the axion with Lagrangian
\begin{align}
    \mathcal{L} 
    &= 
    \frac12 F_a^2 (\partial a)^2
    -\frac{1}{4e^2} F_{\mu\nu} F^{\mu\nu} - \frac{1}{2 g^{2}_{s}} \mathrm{Tr} \left(G_{\mu\nu} G^{\mu\nu} \right)
    \nonumber\\
    &\quad+\sum_{i=1}^{2}  \overline{\Psi}_{i} \left( i\slashed{D} - m_{i} \right) \Psi_{i}  
    + \frac{N a}{8 \pi^{2}} \mathrm{Tr}\left(G_{\mu\nu} \widetilde{G}^{\mu\nu} \right) 
    +  \frac{E a}{16 \pi^{2}} F_{\mu\nu} \widetilde{F}^{\mu\nu},
\end{align}
where $F_{a}$ is the fundamental period of the axion, $E$ is the primordial anomaly of $U(1)_{\sf PQ}$ with $U(1)_{\rm em}$ and $N \in \frac12 \mathbb{Z}$ is the anomaly coefficient of $U(1)_{\sf PQ}$ with QCD.\footnote{Here we have used the unfortunate standard convention making $N$ in general half-integer.}

The condition on $E$ in order for the axion to have $2\pi$ periodicity depends on the the chosen subgroup $\Gamma= 1,\mathbb{Z}_{2},\mathbb{Z}_{3}$ or $\mathbb{Z}_{6}$ of the standard model gauge group $SU(3)\times SU(2)\times U(1)/\Gamma$ \cite{Tong:2017oea}. In this paper we take $\Gamma = \mathbb{Z}_{6}$ in order for the electron to have the minimum quantum of electric charge. With this choice, a sufficient condition for axion $2 \pi$ periodicity is
\begin{equation}
E - \frac{2N}{3} \in \mathbb{Z}  \,.
\label{eq:EandN}
\end{equation}
The axion-gluon coupling explicitly breaks the PQ shift symmetry of the axion. We therefore expect the low-energy effective axion theory to have a potential $V(a)$ and generate an effective axion-photon coupling $g(a)$. The symmetries, phases and domain walls of this theory have been well-studied using the chiral Lagrangian \cite{DiVecchia:2017xpu,Dvali:1996xe,Gabadadze:2000vw,PhysRevD.32.1560} and anomaly matching \cite{Gaiotto:2017tne,Anber:2020xfk}. 

The mass for the axion and its coupling to photons have been calculated at high precision~\cite{GrillidiCortona:2015jxo},
\begin{align}
    m_a^2
    &= 
    \frac{m_{\pi}^{2} f_{\pi}^{2} N^2}{ F_{a}^{2}}
    \left(
    \frac{4}{z + \frac{1}{z}  + 2}
    +\ldots
    \right),
   \qquad 
    g_{a\gamma\gamma}
    =
    \frac{\alpha_{\rm em}}{\pi F_a}
    \left(E - 
    \frac{5}{3}N
    -  \frac{1-z}{1+z} N
    + \ldots
    \right) 
    ,
\end{align}
where $ z = \frac{m_{u}}{m_{d}}$ measures the isospin breaking of $SU(2)_{V}$ and $\ldots$ denote higher order terms in the chiral Lagrangian. 
Note that as is conventional we have written the coupling $g_{a\gamma\gamma}$ in the canonical basis for both axions and photons. 

The usual explanation for the irrational contribution proportional to $\frac{1-z}{1+z}$ is that it arises from the mixing with the pion. This is certainly true, but raises a minor puzzle. In the effective theory, we can integrate out the pion and for all values of the axion, the pion degree of freedom is heavy and the EFT is valid. Therefore, the quantization of the monodromy of the axion-photon coupling should be visible in the effective theory. 

The resolution to this puzzle has been discussed in \cite{Agrawal:2017cmd} and \cite{Fraser:2019ojt} and arises exactly through the monodromic function $g(a)$. As we will show below, the axion coupling to photons can be packaged in this functional form, such that $g(a)$ has integer monodromy under the axion discrete gauge symmetry, but $g'(0)$ can be irrational. 
We review the calculation of the axion potential in the Chiral Lagrangian and derive the form of $g(a)$ relevant for the QCD axion below.  

The effective Lagrangian for the photon, the QCD axion $a$ and the pion $\pi^{0}$, can be written as,
\begin{align}
    \mathcal{L} 
    &= 
    \frac{F_{a}^{2}}{2} \ (\partial a)^{2} + \frac{f_{\pi}^{2}}{2} \ (\partial \pi^{0})^{2} 
    - V(a,\pi^{0})
    +
    {
    \left(E- \frac{5}{3} N \right) \frac{a}{16 \pi^{2}} F\tilde{F}
    +\frac{\pi^0}{16 \pi^{2}} F\tilde{F}
    } \,,
\end{align}
with a potential $V(a,\pi^{0})$ given by
\begin{equation}
V(a,\pi^{0}) = f_{\pi}^{2} m_{\pi}^{2}
    \left(1 - \cos{\frac{2Na}{2}} \cos{\pi^{0}} + 
    \frac{1-z}{1+z}  \sin{\frac{2Na}{2}} \sin{\pi^{0}}
    \right) \,.
    \label{qcdpotential}
\end{equation}
In this basis, the two discrete gauge symmetries involving the axion and the pion are implemented by  $(a,\pi^{0}) \rightarrow (a + 2 \pi,\pi^{0} + 2N\pi)$ and $\pi^{0} \rightarrow \pi^{0} + 2 \pi$. The potential has characteristic eigenvector directions which reverse roles when the sign of $1-z$ flips, which will be important to our discussion throughout.

We would like to study the low-energy 
limit of this theory. In the limit that $f_{\pi} \ll F_{a}$, the pion is much more massive than the axion and can be integrated out. If this can be done consistently at every value of the axion $a$, then axion domain walls are completely describable within the effective field theory. There can in general be additional domain walls (perhaps metastable or unstable) that also involve rearrangements of heavy degrees of freedom or new massless states appearing on the domain walls. These domain walls are not described completely within the EFT. We show that the function $g(a)$ captures this exact behaviour. 

An axion domain wall $a \rightarrow a + 2\pi$ in this particular basis of the potential (equation~\eqref{qcdpotential}) requires a pion domain wall $\pi^{0} \rightarrow \pi^{0} + n \pi$ with $n \in 2N \mathbb{Z}$. For $0 \leq z <1$, the most energetically favourable domain wall is $\pi^{0} \rightarrow \pi^{0} - 2N \pi$ with the $\pi^{0} \rightarrow \pi^{0} + 2N \pi$ domain wall having an additional tension $\Delta T \propto \sqrt{|\frac{1-z}{1+z}|}$. For $1<z<\infty$, the roles of the two domain walls are reversed. 

For any $z \geq 0$, to first order in $\frac{f_{\pi}}{F_{a}}$, we can integrate out the pion using its equation of motion,
\begin{equation}
\frac{\partial V}{\partial \pi^{0}} = 0 \implies \pi^{0} = -\arctan{\left(  \frac{1-z}{1+z} \tan{\frac{2N a}{2}}  \right)} - \pi \mathrm{sign}(1-z) \sum_{k = 1}^{2N} \Theta\left(a-(2k-1)\frac{\pi}{2N} \right)  \,.
\label{pionsolution}
\end{equation}
Here the $\Theta$ Heaviside-function ensures that the axion domain wall is smooth and heavy pion degrees of freedom are not excited. 

This yields the effective Lagrangian as,
\begin{equation}
\mathcal{L}
= \frac{F_{a}^{2}}{2}(\partial a)^{2} -V(a) + 
\frac{g(a)}{16 \pi^2} F_{\mu \nu} \tilde{F}^{\mu \nu},
\label{effectiveaxionpotential}
\end{equation}
with 
\begin{align}
    V(a) &= - f_{\pi}^{2} m_{\pi}^{2}\sqrt{1 - \frac{4 z}{\left(1 + z\right)^2}\sin^{2}\left(\frac{2N a}{2}\right)}  \,, \\ g(a)
    &= E a - 
  \frac{5}{3}N  a- \arctan{\left( \frac{1-z}{1+z}  \tan{\frac{2N a}{2}}  \right)} - \pi \mathrm{sign}(1-z) \sum_{k = 1}^{2N} \Theta\left(a-(2k-1)\frac{\pi}{2N} \right)  \,.
\end{align}
We see the prototypical example of the function $g(a)$ -- it has a monodromy under axion shift symmetry given by
\begin{equation} 
g(a+2\pi) = g(a) + 2\pi \left(E - \frac{2N}{3}  - N \left(1+\mathrm{sign}(1-z)\right) \right)  \,.
\end{equation}
The monodromic charge $\left(E - \frac{2N}{3}  - N \left(1+\mathrm{sign}(1-z) \right) \right) \in \mathbb{Z}$ by equation~\eqref{eq:EandN}. For the specific choice $\frac{E}{N} = \frac{8}{3}$, the monodromy vanishes when $0 \leq z <1$ and is $2N$ when $z > 1$.

The slope around the axion minimum for a generic $z$ is irrational and the axion-photon coupling at this point is given by
\begin{equation}
g_{a \gamma \gamma} = \frac{\alpha_{\mathrm{em}}}{\pi F_{a}} g'(0) = \frac{\alpha_{\mathrm{em}}}{ \pi F_{a}} \left(E - \frac{5}{3} N - \frac{1-z}{1+z} N \right)  \,.
\end{equation}
Note that both the potential $V(a)$ and $g(a)$ depend on the axion-pion mixing parameter $z$ and that $g(a)$ is quantized exactly in the limit $z \rightarrow \{0,\infty\}$ when the mass vanishes. Under the transformation $z \leftrightarrow \frac{1}{z}$, the effective potential is left unaltered and the monodromy of $g(a)$ changes by $2N$.

In the isospin restoring limit $z \rightarrow 1$, the potential (equation~\eqref{qcdpotential}) has an additional $\mathbb{Z}_{2} \subset SU(2)_{V}$ pion parity $(-1)^{N_{\pi}}$ symmetry that sends $\pi^{0} \rightarrow - \pi^{0}$ and the tension difference between the domain walls $\pi^{0} \rightarrow \pi^{0} \pm 2N \pi$ goes to zero. The profiles and monodromies for $\lim\limits_{\ {z\to 1^-}}g(a)$ and $\lim\limits_{\ z\to1^+} g(a)$ differ as this $\mathbb{Z}_{2}$ is broken by the Wess-Zumino-Witten term. Additionally, in the limit $z \rightarrow 1$, the pion shift symmetry is restored at $a = \pi$ and the pion becomes massless. The potential $V(a)$ has a corresponding cusp at this point and $g(a)$ is discontinuous due to the massless pion jump. This cusp and the discontinuity at $a=\pi$ is an accidental restoration of the pion shift symmetry at $a = \frac{\pi}{2N}$ for $N_{f} =2$ at this order in the chiral Lagrangian and is resolved at higher orders~\cite{Smilga:1998dh}. 

In the next few sections, we shall see several physical systems with the same prototypical $g(a)$ but different potentials $V(a)$.

\section{Perturbative PQ breaking} \label{PQbreaking}
It is very instructive to compare the breaking of the axion shift symmetry by QCD effects to a perturbative form of shift symmetry breaking. The simplest such model is a massive charged Dirac fermion $\Psi$ coupled to a $U(1)$ gauge field and an axion $a$ coupled through a chiral mass term, resulting in a Lagrangian of the form
\begin{equation}
\mathcal{L} = i\overline{\Psi} \slashed D \Psi - f \overline{\Psi} e^{i a \gamma_{5}} \Psi - m_{\Psi} \overline{\Psi} \Psi  \,.
\end{equation}
This Lagrangian has an axion shift symmetry $a \rightarrow a + c$ and $\Psi \rightarrow e^{-i \frac{c}{2}\gamma_{5}} \Psi$, which is explicitly broken by the mass term $ m_{\Psi}$. We therefore expect to generate both a $g(a)$ and $V(a)$ and the low-energy effective Lagrangian should be of the form equation~\eqref{lowenergyeft}. We shall see that this simple model captures a lot of the features of the QCD axion. 

\subsection{The effective potential}
In order to calculate $g(a)$ and $V(a)$ in this simple model, we compute the effective action by integrating out the massive fermion,
\begin{equation}
iS_{\mathrm{eff}} = \mathrm{Tr}\ln{ \left[i\left(i\slashed D -  m_{\Psi} - f e^{i a \gamma_{5}}\right) \right]}  \,.
\label{pqefa}
\end{equation}
This yields an effective potential for a constant axion $a$ as
\begin{equation}
V(a) = 2i\mathrm{Tr}\ln{\left[\partial^{2} +  m_{\Psi}^{2} + 2  m_{\Psi} f \cos{a} + f^{2}\right]}  \,.
\end{equation}
This is simply the Coleman-Weinberg potential following from a particle with an effective mass
\begin{equation}
m(a)^{2} =  m_{\Psi}^{2}  + 2  m_{\Psi} f \cos{a} + f^{2} = ( m_{\Psi}+f)^{2} \left(1- \frac{4}{ \frac{1}{z}+ z + 2} \sin^{2} \left(\frac{a}{2}\right) \right)  \,,
\end{equation}
where we have defined the parameter
\begin{equation}
z = \frac{f}{ m_{\Psi}}  \,.
\end{equation}
The Coleman-Weinberg potential $V(a)$ is
\begin{equation}
V(a) = -c_{1} m(a)^{2} - \frac{m(a)^{4} }{16 \pi^{2}} \ln{\frac{m(a)^{2}}{c_{2}^{2}}}  \,,
\end{equation}
where $c_{1}$ and $c_{2}$ are renormalization scheme-dependent quantities.

This potential shares many features of the QCD axion. For instance, the potential becomes axion independent when $z \rightarrow \{0,\infty\}$ as this is when either $f$ or $m_{\Psi}$ are zero, the latter being the shift symmetry restoring limit in which the axion can be rotated into a $F \widetilde{F}$ term. 

In the limit $z=1$, the effective mass of the particle $m(a)$ vanishes at the chiral symmetry restoring point $a = \pi$ and should not be integrated out. This is reflected by a singularity in $V''''$ at $a = \pi$. 

\subsection{The effective axion-photon coupling}
Just as with the QCD axion, one expects to generate an effective axion-photon coupling $g(a)$ from shift symmetry breaking. Such terms can be calculated by first taking a derivative of the effective action (equation~\eqref{pqefa}) with respect to $a$ to obtain
\begin{equation}
\frac{\delta S_{\mathrm{eff}}}{\delta a} = -\mathrm{Tr}\left(\frac{\gamma_{5}f e^{ia \gamma_{5}}}{i\slashed D -  m_{\Psi} -f e^{i a \gamma_{5}}}\right) =  \mathrm{Tr}\left(\frac{\gamma_{5} f(i e^{ia \gamma_{5}}\slashed D  +  m_{\Psi} e^{ia \gamma_{5}} + f)}{ \slashed D^{2} + m(a)^{2}}\right)  \,.
\label{PQbreakingeffaxion}
\end{equation}
The trace is both over spinor indices as well as the implicit momentum integrals and this time we keep both a constant axion $a$ and a constant field strength $F_{\mu \nu}$, such that $\left(\slashed  D\right)^{2} = D^{2} - \frac{1}{2}\sigma_{\mu \nu} F^{\mu \nu}$ with $\sigma_{\mu \nu} = \frac{i}{2}[\gamma_{\mu},\gamma_{\nu}]$. By matching equation \eqref{PQbreakingeffaxion} with the same derivative of the low-energy effective axion-Maxwell action (Eq. \eqref{lowenergyeft}), we find $g'(a)$. 

Since $g'(a)$ is CP even, we expand equation \eqref{PQbreakingeffaxion} to second order in $F$ and keep only CP odd terms as
\begin{equation}
\frac{\delta S_{\mathrm{eff}}}{\delta a}|_{\mathrm{odd}} =  \frac{1}{4}\mathrm{Tr}\left( \gamma_{5} \frac{ f^{2}+  m_{\Psi}f \cos{a}}{\left( D^{2} + m(a)^{2}\right)^{3}} \sigma^{\mu \nu} F_{\mu \nu} \sigma^{\alpha \beta} F_{\alpha \beta}\right)  \,.
\end{equation}
The trace can now be reduced by using the gamma matrix identity $\mathrm{Tr}\left(\gamma^{5} \sigma_{\mu \nu} \sigma_{\alpha \beta}\right)= -i 4\epsilon_{\mu \nu \alpha \beta}$ and inserting a trace over four momenta yields
\begin{equation}
    \frac{\delta S_{\mathrm{eff}}}{\delta a}|_{\mathrm{odd}} =  2i \int \frac{d^{4}p}{(2\pi)^{4}} \frac{ f^{2}+  m_{\Psi}f \cos{a}}{\left( p^{2} - m(a)^{2}\right)^{3}} F \widetilde{F}  \,.
\end{equation}
The momentum integrals are convergent and rewriting in terms of the order parameter $z$ yields
\begin{equation}
\frac{\delta S_{\mathrm{eff}}}{\delta a}|_{\mathrm{odd}} =  \frac{1}{16 \pi^{2}} \left(\frac{z^{2} + z \cos{a}}{1 +  2 z \cos{a} + z^2} \right) F \widetilde{F}  \,.
\end{equation}
Comparing this result to the effective low-energy action of the axion (equation~\eqref{lowenergyeft}) allows for the matching
\begin{equation}
g'(a) =  \frac{z^{2} + z \cos{a}}{1 +  2 z \cos{a} + z^2}   \,.
\end{equation}
Integrating this function yields the effective low energy axion-photon coupling as
\begin{equation}
g(a) = \frac{1}{2} a - \mathrm{arctan}\left( \frac{1-z}{1+z} \tan{\frac{a}{2}} \right) - \mathrm{sign}(1-z) \pi \Theta(a-\pi)  \,.
\end{equation}
Similar to the potential, $g(a)$ captures many features of the symmetries of the Lagrangian in its fully summed form. For instance, in the limit $z= \infty$ or $ m_{\Psi} =0$, $g(a)$ becomes $a F \widetilde{F}$. When $z = 0$, the effective coupling to photons vanishes as $f = 0$. Thus in both cases $g(a)$ becomes of the form $\mathbb{Z}a$  when the axion becomes massless as predicted by general symmetry arguments. 

Similar to the QCD axion, in the limit $z \rightarrow 1$, there is an apparent restoration of a $\mathbb{Z}_{2}$ symmetry that acts 
on the fields as
\begin{equation}
\Psi(t,x) \rightarrow \gamma^{0} \Psi(t,-x) \qquad A_{\mu}(t,x) \rightarrow (-1)^{\mu} A_{\mu}(t,-x) \qquad a(t,x) \rightarrow a(t,-x)  \,,
\end{equation}
where $(-1)^{\mu} = 1$ if $\mu = t$ and $-1$ otherwise. This symmetry leaves the Lagrangian invariant in this limit up to the change of the Chern-Pontryagin density $F \widetilde{F}$. Correspondingly, the profiles for $\lim\limits_{\ {z\to 1^-}}g(a)$ and $\lim\limits_{\ z\to1^+} g(a)$ differ and the monodromies are respectively $0$ and $1$. Both profiles of $g(a)$ also have a discontinuous jump at the chiral symmetry restoring point $a = \pi$ where the fermion becomes massless.

\section{Axions from extra dimensions and $g(a)$}\label{sec:5dfermions}
A class of interesting axions is those that descend from gauge theories and higher form fields in extra dimensions. These are particularly motivated both due to the fact that they arise generically in string compactifications as well as due to high quality global symmetry that descends from a gauge symmetry in the bulk.   

In these models, axion potentials  arise from charged objects wrapping internal cycles, which appear as instantons in the $4$D theory, see e.g. \cite{Dine:1986zy,Dine:1987bq,Becker:1995kb,Witten:1996bn,Ooguri:1996me}. Alternatively, this potential can be thought of as arising from the axion dependence of a KK tower of states which undergoes spectral flow as axion $a \rightarrow a + 2\pi$. A similar effect arises from a tower of dyonic states in axion-Maxwell theory \cite{Fan:2021ntg}.    

In this section we bridge the relation between these two sources for the axion potential through an instructive example. In doing so, we show that massive charged fermions with additional compact degrees of freedom coupled to the axion generate a similar axion potential $V(a)$ and an effective axion-photon coupling $g(a)$. 

In appendix \ref{appendix2}, we reformulate the results of this section in terms of an effective worldline formalism of a  chargded massive $4$D fermion with additional compact degrees of freedom coupled to the axion. In doing so, we derive the effective axion-Euler-Heisenberg Lagrangian to all orders in constant $a$ and $F$.

\subsection{5D instantons}
We consider a $U(1)$ gauge theory with gauge field $A$ in  5D Euclidean space ($g_{\mu \nu} = \delta_{\mu \nu}$) with a massive charged fermion $\Psi$, with the fifth dimension $y$ compactified on a circle of radius $R$,
\begin{align}
    S
    &=
    \int d^4x \int dy\,
    \left[
    -\frac{1}{4e^{2}} F_{MN} F^{MN}
    - \Psi^{\dagger} \left(\slashed{D} + m 
 \right)\Psi
    \right]  \,.
    \label{5dlagrangian}
\end{align}
 The axion is identified with a Wilson loop $\int dy A$ around the compact extra dimension in almost axial gauge as
\begin{equation}
A_{5}(x,y) = \frac{a(x)}{2\pi R}  \,.
\end{equation}
In any theory with a compact dimension, the modes of the particle can be understood in terms of a tower of states (KK modes). In the present theory, this leads to a description of the $5$D fermion as a tower of electrically charged massive $4$D fermions with axion-dependent masses.

An alternative and more useful representation for our purposes is in terms of winding modes of the fermion around the compact dimension. Non-local loops of the fermion around the compact dimension appear as instanton effects in $4$D, giving a mass to the axion. In such a formulation, the axion dependence of the theory can be put into a twisted boundary condition for the fermions~\cite{Arkani-Hamed:2007ryu},
\begin{align}
    \Psi(y+2\pi n R)
    &\simeq
    e^{in (\pi-a)} \Psi (y)  \,,
\end{align}
in which we have also given the fermion additional anti-periodic boundary conditions to align the minimum of the potential with $a = 0$.

 The Green's function on the compactified space $G$ in the presence of an axion can similarly be decomposed as a sum over twisted flat space Green's functions $D_{F}$ as
\begin{equation}
    G(x,y) = \sum_{n = -\infty}^{\infty} e^{in (a+\pi)}  D_{F}(x,y+2\pi n R)  \,.
    \label{compact to flat}
\end{equation}
Thus, only fermion propagators that loop around the extra dimensions are sensitive to an axion background, and can hence generate a potential for the axion and an effective axion-photon $g(a)$\footnote{On flat space, integrating out $5D$ fermions generates a level $\frac{1}{2}$-Chern-Simons term. In order to still have an axion gauge symmetry, the action therefore implicitly includes a primordial level $\pm \frac{1}{2}$ Chern-Simons term, which won't concern us for the rest of the calculation. \label{foonote1}}.
This effect is suppressed by the small spacelike propagator for heavy $\Psi$ to loop around the extra dimension  $z =  e^{- 2 \pi R m}$, such that the instanton contributions to the axion effective action can be packaged in the complex number quantity,
\begin{equation}
 \mathcal{Z} = e^{- 2 \pi R m } e^{ia}  \,.
\end{equation}
The real and imaginary parts of powers of $\mathcal{Z}$ respectively contribute to the CP even potential $V(a)$ and CP odd effective axion-photon coupling $g(a)$. Similar contributions were noted to the superpotential in \cite{Dine:1987bq}. 

The equivalence between a tower of states (e.g. dyons) and instantons is exactly given by the equivalence between the KK mode and winding modes formulation \cite{Cheng:2002iz} (of which equation \eqref{compact to flat} is a special example). The relation between the two is provided by Poisson resummation \cite{Fan:2021ntg},
\begin{equation}
\sum_{n = - \infty}^{\infty} s \left(n - \frac{a}{2 \pi } \right) = \sum_{k = - \infty}^{\infty} e^{-i k a} S(k)  \qquad , \qquad S(k) = \int_{-\infty}^{\infty}dx \ e^{-i 2\pi k x} 
s(x)   \,.
\end{equation}


We proceed to calculate both the potential $V(a)$ and effective axion-photon coupling $g(a)$ by evaluating the effective action (Eq. \eqref{5dlagrangian}) after integrating out the massive charged fermions in the winding mode basis. An alternative derivation of both $V(a)$ and $g(a)$ using KK modes can be found in the appendix \ref{appendix1}. 

\subsection{Calculation of $V(a)$}
In order to obtain an effective action for the axion and photons, we integrate out the fermions to obtain
\begin{equation}
e^{ S_{\rm eff}[a,A]} = \int D\Psi D \overline{\Psi}\ e^{ S[a,A,\Psi]}  \,.
\end{equation}
This yields an effective action
\begin{equation}
S_{\rm eff}[a,A] = S[a,A] + \mathrm{Tr}\left(  \log{\left(  -\slashed D  - m   \right)}  \right)  \,.
\label{effectiveaction}
\end{equation}

A simple way to calculate $V(a)$ is to take a derivative of the effective action \eqref{effectiveaction} with respect to a constant axion $a$ and set the photon field to zero. This yields
\begin{equation}
\left(2 \pi R \right) \frac{\delta S_{\mathrm{eff}}}{\delta a} \supset -\mathrm{Tr}\left(\gamma_{5}G \right)  \,.
\label{eq:efa5d}
\end{equation}
The compact Green's function $G$ can be expanded in terms of twisted flat space Green's functions as
\begin{equation}
\mathrm{Tr} \left(\gamma_{5}G \right) = \sum_{n = - \infty}^{\infty} e^{in (a+\pi)}  \mathrm{Tr}\left(i\gamma_{5}\frac{  e^{n 2\pi R \partial_{5}}}{\slashed  D + m }\right)  \,.
\end{equation}
Multiplying top and bottom by the same factor, one arrives at
\begin{equation}
\mathrm{Tr} \left(\gamma_{5}G \right) = \sum_{n = - \infty}^{\infty} e^{in (a+\pi)} \mathrm{Tr}\left(i\gamma_{5} \frac{ e^{ n 2\pi R \partial_{5}} (\slashed D - m)}{\left(\slashed  D\right)^{2} - m^{2} }\right)  \,.
\label{dsda}
\end{equation}
This allows us to calculate the $4$D potential by equation \eqref{eq:efa5d} as
\begin{equation}
\frac{\partial V}{\partial a} = 4 \sum_{n = - \infty}^{\infty}  e^{in (a+\pi)}  \int \frac{dp^{5}}{(2\pi)^{5}} \frac{ \ p_{5}e^{i n 2\pi R p_{5}} }{p^{2}+(p_{5})^{2} + m^{2}}  \,.
\end{equation}
Note that this average of momentum in 5D is non-zero due to the discrete nature of the momenta. Integrating these out\footnote{The $n=0$ term is axion independent and vanishes due to CP symmetry.} yields
\begin{equation}
\frac{\partial V}{\partial a} = -4\sum_{n=1}^{\infty} (-1)^{n}\sin{\left(na\right)} \int \frac{dp^{4}}{(2\pi)^{4}}  e^{- n|  2 \pi R |  \sqrt{p^{2}+m^{2}}}  \,.
\end{equation}
Integrating over momenta and with respect to the axion $a$ yields the effective potential
\begin{equation}
V(a) = \frac{m^{2}}{(2 \pi R)^{2}}\sum_{n=1}^{\infty} \frac{1}{ \pi^{2} n^{3}} e^{-n 2 \pi |R m|}  (-1)^{n} \cos{\left(na\right)}\left( 1 + \frac{3}{2 \pi |R m| n} + \frac{3}{\left(2 \pi R m n\right)^{2}} \right)  \,.  
\label{5dpotential}
\end{equation}
This potential is known as the one generated by a four-dimensional particle with a rotor degree of freedom coupled to the axion \cite{Fan:2021ntg} and was also discussed in various limits in \cite{Arkani-Hamed:2007ryu,Sundrum:2005jf,Cheng:2002iz,Arkani-Hamed:2003xts}. Various other representations of this potential are recorded in the appendices in equation~\eqref{wordlinepotential} and equation~\eqref{KKpotential}.

We see that for the 5D instantons, the spurion is parametrized by the parameter $z = e^{-2 \pi R m}$ with the symmetry $z \leftrightarrow \frac{1}{z}$ leaving the potential invariant and implementing the $-m$ to $m$ domain wall. 

At the symmetric point $z = 1$, the 5D fermion becomes massless and the Lagrangian has an apparent $\mathbb{Z}_{2}$ ($5$D parity) symmetry, which is broken by the topological Chern-Simons term. In this limit, at the point $a = \pi$, the lightest $4$D fermion in the tower becomes massless and a $4$D $U(1)$-chiral symmetry is restored, meaning that this fermion should not have been integrated out. This is reflected by a singularity of $V''$ at $a = \pi$.

\subsection{Calculation of $g(a)$}
In this section, we calculate the effective axion-photon coupling resulting from a charged massive  fermion with additional compact degrees of freedom coupled to the axion. The existence of such a $g(a)$ was already well-known in the context of finite temperature field theory in various dimensions see e.g. \cite{Dunne:1996yb,Sisakian:1997cp} and references therein. 

We can calculate $g(a)$, the effective axion-photon coupling, by taking a derivative of the effective action (equation \eqref{effectiveaction}) with respect to the axion, which will contain a term of the form
\begin{equation}
i \frac{g'(a)}{16 \pi^{2}} F \tilde{F} \subset(2\pi R) \frac{\delta  S_{\mathrm{eff}}}{\delta a}  \,.
\label{gofa5d}
\end{equation}
The calculation of $g(a)$ will proceed along similar lines as in section \ref{PQbreaking}. We keep both a background constant axion $a$ and a constant background field of the zero KK mode of the photon $F^{(0)}_{\mu \nu}$ which we simply write as $F_{\mu \nu}$. In these circumstances, $\left(\slashed  D\right)^{2} = D^{2} - \frac{1}{2}\sigma_{\mu \nu} F^{\mu \nu}$.

Returning to the effective action in equation \eqref{eq:efa5d} and keeping only the relevant CP odd terms as
\begin{equation}
(2\pi R)\frac{\delta  S_{\mathrm{eff}}}{\delta a}|_{ \mathrm{odd}} = mi\sum_{n = - \infty}^{\infty} e^{in (a+\pi)} \mathrm{Tr}\left(\gamma_{5} \frac{ \ e^{n 2\pi R \partial_{5}} }{\left(\slashed  D\right)^{2} - m^{2} }\right)  \,.
\end{equation}
Expanding this to second order in $F$, one obtains
\begin{equation}
(2\pi R)\frac{\delta  S_{\mathrm{eff}}}{\delta a}|_{ \mathrm{odd}} =  i\frac{m}{4}\sum_{n = - \infty}^{\infty} e^{in (a+\pi)} \mathrm{Tr}\left(\gamma_{5} \frac{ \ e^{n 2\pi R \partial_{5}} }{D^{2} - m^{2} } 
\sigma^{\mu \nu} F_{\mu \nu} \sigma^{\alpha \beta} F_{\alpha \beta} \right)  \,.
\end{equation}
Using the identity $\mathrm{Tr}\left(\gamma^{5} \sigma_{\mu \nu} \sigma_{\alpha \beta}\right)= -4 \epsilon_{\mu \nu \alpha \beta}$ one arrives at
\begin{equation}
(2\pi R)\frac{\delta  S_{\mathrm{eff}}}{\delta a}|_{ \mathrm{odd}}  = 2mi \sum_{n = - \infty}^{\infty} e^{in (a+\pi)} \int \frac{dp^{5}}{(2\pi)^{5}} \frac{ \ e^{i n 2\pi R p^{5}} }{\left((p^{5})^{2} + p^{2} + m^{2} \right)^{3} } F \widetilde{F}  \,.
\end{equation}
The integrals over momenta are convergent. Performing the integrals yields
\begin{equation}
(2\pi R) \frac{\delta  S_{\mathrm{eff}}}{\delta a}|_{ \mathrm{odd}}  = \frac{i}{32 \pi^{2}} \mathrm{sign}(m)\sum_{n = - \infty}^{\infty} e^{in (a+\pi)} e^{- 2 \pi |R m n|} F \widetilde{F}  \,.
\end{equation}
By comparing with equation \eqref{gofa5d}, we find that
\begin{equation}
g'(a) =  \frac{\mathrm{sign}(m)}{2}\sum_{n = - \infty}^{\infty} e^{in (a+\pi)}e^{- 2 \pi |R m n|}  \,.
\end{equation}
This sum can be explicitly calculated and yields a $g'(a)$ of the form
\begin{equation}
g'(a) = \frac{1}{2}\frac{\sinh{2\pi R m}}{\cosh{2\pi R m} + \cos{a}}  \,.
\end{equation}
Integrating this with respect to $a$ yields our final result for $g(a)$ and adding a primordial $\pm \frac{1}{2}$-level Chern-Simons term (see foonote \ref{foonote1}) yields
\begin{equation}
g(a) =  \pm \frac{1}{2}a + \arctan{\left(\frac{1-z}{1+z} 
 \tan{\left( \frac{a}{2}\right)} \right)} + \pi \mathrm{sign}(1-z) \Theta (a-\pi)  \,.
\label{5dga}
\end{equation}
In the limit $R \rightarrow \infty$, this reproduces the well-known result,
\begin{equation}
g(a) = \frac{1}{2}\left(\pm 1 + \frac{m}{|m|} \right) a  \,.
\end{equation}
Similar to the QCD axion, there is an apparent $\mathbb{Z}_{2}$ ($5$D parity) restoration as $m \rightarrow 0$ or $z \rightarrow 1$ in equation \eqref{5dlagrangian}. The profiles and monodromies of $\lim\limits_{\ {z\to 1^-}}g(a)$ and $\lim\limits_{\ z\to1^+} g(a)$ differ however due to the gauge-parity anomaly. In the same limit $z \rightarrow 1$, there is a jump in both profiles of $g(a)$ at $a = \pi$ due to the lightest fermion in the tower becoming massless. 

In the 5D theory, a domain wall describing the $-m \rightarrow m$ transition has a massless chiral fermion on it and describes anomaly inflow consistent with our $4$D analysis.

\section{Discussion}
We have considered the general properties of the monodromic axion-photon coupling $g(a)$ and the symmetries of the low-energy axion-Mawell Lagrangian in the presence of such a coupling. We argued that the non-quantization of $g'(a)$ is a spurion for the axion shift symmetry. The connection between the axion potential and this coupling has been considered supported by several examples including the QCD axion, perturbative shift symmetry breaking and fermions with additional compact degrees of freedom. In all such cases, a prototypical monodromic function $g(a)$ was derived and could be expanded in terms of a linear monodromic function with same monodromic charge as $g(a)$ and a periodic function. In some cases only the first few terms in the expansion of the periodic function dominated. However, in many simple models the whole Fourier tower needs to be summed up to get the relevant $g(a)$. In such cases $g(a)$ captured the rearrangement of heavy degrees of freedom through its singularities at isolate points. This correlated with cusps and singularities in the axion potential.

There are a number of model building applications of this formulation. Instead of building effective field theories with polynomial axion couplings, more general non-linear couplings can arise naturally through the $g(a)$ portal. This may have interesting avenues for constructing more general natural potentials for axions. Phenomenologically, the full non-linear form of $g(a)$ is most relevant for scenarios where the axion traverses an $O(1)$ fraction of its field range. This is certainly true for axion strings and domain walls, and  sharp features in $g(a)$ can affect axion emission from these objects. It can also be true for dense axion objects, like axion miniclusters or superradiant axion clouds surrounding rotating black holes. 


It will be interesting to study the effective photon coupling for mesons in the chiral Lagrangian, e.g.~for the pion $g(\pi^{0})$ after integrating out $\eta'$. It has been shown~\cite{Komargodski:2018odf}  in the context of a one-flavor QCD $N$ that degrees of freedom rearrange on the $\eta' \rightarrow \eta' + 2 \pi$ domain wall, leading to a fractional quantum hall droplet  and a potential jump in $g(\eta')$. It will be nice to see this physics captured within the effective field theory.  

Lastly, several contributions to $g(a)$ could be considered in the presence of CP-odd sources of axion shift symmetry breaking such as magnetic monopoles (with fermions) and non-Abelian instantons. \\

\noindent \emph{Note added}: While this manuscript was being finalized we became aware of other studies appearing today~\cite{Reece:2023iqn,Choi:2023pdp} which also consider quantization of the axion-gauge couplings. The main focus of these works is different from the non-linear coupling to photons highlighted in this paper.

\acknowledgments
We would like to thank Mario Reig for useful comments on the draft and John March-Russell, Michael Nee and Thomas Harvey for helpful discussions. PA is supported by the STFC
under Grant No. ST/T000864/1. AP is supported by a STFC Studenship No. 2397217 and Prins Bernhard Cultuurfondsbeurs No. 40038041 made possible by the Pieter Beijer fonds and the Data-Piet fonds.

\appendix
\numberwithin{equation}{section}
\section{Axion-Euler-Heisenberg}\label{appendix2}
In this appendix we connect our results back to those of a $4$D worldline particle with a compact additional degrees of freedom coupled to the axion. We do this using the Schwinger proper time formalism and consider a fermionic particle with electrically charged translational modes $x^{\mu}$ and a compact additional degree of freedom $q$ which is coupled to the axion $a$. In doing so, we show how a potential $V(a)$ (equation~\eqref{5dpotential}) and effective axion-photon coupling $g(a)$ (equation~\eqref{5dga}) arise in such a wordline formalism. This will provide us with the low energy effective axion Maxwell field theory to all orders in $a$ and $F$ by studying the axion-Euler-Heisenberg Lagrangian resulting from integrating out such fermions. Axial couplings to the worldlines of particles have been studied in \cite{Mondragon:1995ab,DHoker:1995aat}. The effective Euler-Heisenberg Lagrangian following from loops of fermions with couplings to non compact psuedoscalar particles has been studied in \cite{Jacobson:2018kso}.

We can derive the $4$D wordline formalism of a fermion with additional compact degrees of freedom coupled to the axion by starting from the $5$D effective action \eqref{effectiveaction}, repeated here for completeness,
\begin{equation}
S \supset  \mathrm{Tr}\left( \ln\left(-\slashed D - m \right)\right)  \,.
\label{sec5:effaction}
\end{equation}
The presence of a psuedoscalar (the axion) implies that the Euclidean effective action has both a real and imaginary part as the operator $\slashed D$ no longer has a positive definite spectrum. For this reason, the contributions to the effective action are split into a real and imaginary part as
\begin{equation}
S \supset  \mathrm{Tr}\left( \ln \left|\slashed D + m \right|\right) + i  \mathrm{Tr}\left( \mathrm{Arg}\left( -\slashed D - m \right) \right)  \,.
\label{realandimag}
\end{equation}
For our purposes, this split is done by taking a derivative of the effective action \eqref{sec5:effaction} with respect to $m^{2}$ as
\begin{equation}
\frac{dS}{dm^{2}} \supset \frac{1}{2} \mathrm{Tr}\left( \frac{1}{-\slashed D^{2} + m^{2}}\right) - \frac{1}{2m} \mathrm{Tr}\left(\frac{\slashed D}{-\slashed D^{2} + m^{2}} \right)  \,.
\label{wordline}
\end{equation}
The first term in equation \eqref{wordline} can be reformulated using standard techniques \cite{PhysRev.80.440,Corradini:2015tik}  in terms of a wordline effective action for a fermion with $4$ translation degrees of freedom $x^{\mu}$ and one additional compact degree of freedom $q$ as
\begin{equation}
S \supset   -\frac{1}{2} \int_{0}^{\infty} \frac{ds}{s} e^{-sm^{2}} \int  Dq e^{- \int_{0}^{s} d\tau \left(\frac{1}{4} \dot{q}^{2} - i \dot{q} \frac{a}{2\pi} \right)} \int Dx e^{- \int_{0}^{s} d\tau \left(\frac{1}{4} \dot{x}^{2} - i \dot{x}_{\mu} A^{\mu} \right)} \mathrm{Spin}[x, A]  \,.
\end{equation}
Here we have isolated the compact fifth degree of freedom $q$ of the fermion that is coupled to the axion and the spin factor is given by
\begin{equation}
\mathrm{Spin}[x, A] =  \mathrm{Tr} P \mathrm{exp}\left[-\frac{i}{4} [\gamma^{\mu},\gamma^{\nu}] \int_{0}^{s} d\tau F_{\mu \nu}(x(\tau)) \right]  \,.
\end{equation}
From the effective action it is clear that this term will generate an effective potential $V(a)$ for the axion $a$ and an effective Euler-Heisenberg Lagrangian, which shall be calculated in the next section. This effective potential was studied for loops of bosonic particles with additional rotor degree of freedoms coupled to the axion in reference~\cite{Fan:2021ntg}.

Importantly however, the second term  in equation~\eqref{wordline} does not vanish when an axial coupling is present. This term generates an effective axion photon coupling. We thus see that loops of fermionic particles with an additional compact degree of freedom $q$ and charged translation modes $x^{\mu}$ can generate an effective $F \widetilde{F}$ coupling when this additional degree of freedom is coupled to the axion. 

We now proceed to calculate both the real and imaginary contributions (Eq. \eqref{realandimag}) to the effective action resulting in an effective axion-Euler-Heisenberg Lagrangian. We shall do this calculation in terms of the KK mode decomposition with frequencies $\omega_{n} = \frac{\pi}{2 \pi R}(2n+1)$ and $n \in \mathbb{Z}$.

The first term in equation \eqref{wordline} can be rewritten using Schwinger-proper time as
\begin{equation}
\frac{1}{2} \mathrm{Tr}\left( \frac{1}{-\slashed D^{2} + m^{2}}\right) = \frac{1}{2} \int_{0}^{\infty}ds \ e^{-sm^{2}} \mathrm{Tr} \left[ \braket{x|e^{s \slashed D^{2}}|x}\right]  \,.
\label{firstcontr}
\end{equation}
We proceed by splitting the covariant derivative $(\slashed D)^{2} = (\slashed D_{4})^{2} + (\slashed D_{5})^{2}$ into the $4$D covariant derivative $\slashed D_{4}$ and the covariant derivative over the fifth dimension $\slashed D_{5}$ and have taken the axion $a$ and field strength $F$ to be constant. 

The trace over the $4$D covariant derivative in the presence of a constant field strength $F$ can be calculated using the well-known identity in $4$D \cite{Dunne:2004nc,Heisenberg:1936nmg}, 
\begin{equation}
\mathrm{Tr} \left[\braket{x|e^{\slashed D^{2}_{4}s}|x} \right] = \frac{1}{64 \pi^{2}} \frac{F \widetilde{F}}{\mathrm{Im} \cosh{\left( sX\right)}} \mathrm{Tr} \left[\mathrm{exp}\left(-  \frac{s}{2} \sigma_{\mu \nu} F^{\mu \nu} \right) \right]   \,,
\label{schwingercalc}
\end{equation}
and the trace identity as
\begin{equation}
    \mathrm{Tr} \left[\mathrm{exp}\left(-  \frac{s}{2} \sigma_{\mu \nu} F^{\mu \nu} \right)\right] = 4 \ \mathrm{Re} \cosh{sX}  \,,
\end{equation}
with $X$ given by
\begin{equation}
    X \equiv \sqrt{\frac{1}{2} F^{2} + \frac{i}{2} F \widetilde{F}}  \,.
\end{equation}
We can now calculate the action $S$ by plugging these identities into equation~\eqref{firstcontr} and integrating with respect to $m^{2}$ (equation~\eqref{wordline}) to obtain the well-known formula for the Euler-Heisenberg Lagrangian. In case of a constant axion $a$ and field strength $F_{\mu \nu}$ this is
\begin{equation}
     \mathcal{L} \supset  -\frac{1}{32 \pi^{2}} \int_{0}^{\infty} \frac{ds}{s} e^{-s m^{2}} \frac{\mathrm{Re}\cosh{sX}}{\mathrm{Im}\cosh{sX}}  F \widetilde{F}  \frac{1}{\left(2 \pi R \right)} \sum_{n = - \infty}^{\infty} e^{- \left(\omega_{n} - \frac{a}{2 \pi R} \right)^{2} s}  \,.
    \label{piece1}
\end{equation}
By expanding this formula in powers of $s$, we can find an alternative integral representation for the potential of the axion. Observe that to second order in $s$,
\begin{equation}
\frac{\mathrm{Re}\cosh{sX}}{\mathrm{Im}\cosh{sX}}  F \widetilde{F} = \frac{4}{s^{2}} + \frac{2}{3} F^{2} + \mathcal{O}(s^{2})  \,.
\end{equation}
We recognize the first term as the vacuum energy contribution to the Euler-Heisenberg Lagrangian. This yields an alternative integral representation for the potential of the axion of the form
\begin{equation}
V(a) = \frac{1}{8 \pi^{2}} \int_{0}^{\infty} \frac{ds}{s^{3}} e^{-sm^{2}}  \sum_{n = - \infty}^{\infty} e^{- \left(\omega_{n} - \frac{a}{2 \pi R} \right)^{2} s}  \,,
\label{wordlinepotential}
\end{equation}
and can be rewritten in terms of instanton supressed contributions using Poisson resummation \cite{Fan:2021ntg}. 

We now proceed to calculate the imaginary part of the effective action (equation~\eqref{realandimag}) by calculating the contribution of the second term in equation~\eqref{wordline}. In doing so, we recover an alternative representation for $g(a)$ and complete the axion-Euler-Heisenberg Lagrangian.

The second term in equation \eqref{wordline} can be calculated in a similar manner using Schwinger proper time
\begin{equation}
-\frac{1}{2m} \mathrm{Tr}\left(\frac{\slashed D}{-\slashed D^{2} + m^{2}} \right) = -\frac{1}{2m}\int_{0}^{\infty} ds \ e^{- s m^{2}} \mathrm{Tr} \left[\braket{x|\gamma_{5} \left( \partial_{5} - i \frac{a}{2 \pi R}\right)e^{  \slashed D^{2} s}|x}\right]  \,.
\end{equation}
This term is non-zero due to the discrete nature of the momenta of the additional degree of freedom. 

This term can now be straightforwardly calculated by again splitting the covariant derivative and using $(\slashed D_{4})^{2} = D^{2} - \frac{1}{2} \sigma_{\mu \nu}F^{\mu \nu}$ and the trace identity identity
\begin{equation}
 \mathrm{Tr}\left[\gamma_{5} \mathrm{exp}\left(-  \frac{s}{2} \sigma_{\mu \nu} F^{\mu \nu} \right)\right] =  -4 \  \mathrm{Im} \cosh{sX}  \,.
\end{equation}
Using this latter trace identity and our expression for the $4$D propagator (equation~\eqref{schwingercalc}), we can calculate the imaginary contribution to the effective action as
\begin{equation}
\frac{d \mathcal{L}}{dm^{2}} \supset \frac{i}{32 \pi^{2} m}\int_{0}^{\infty} ds  \ e^{-s m^{2}} F_{\mu \nu} \tilde{F}_{\mu \nu} \sum_{n = - \infty}^{\infty} \left(\omega_{n} - \frac{a}{2 \pi R} \right) e^{- \left(\omega_{n}-\frac{a}{2 \pi R} \right)^{2} s}  \,.
\end{equation}
Performing the integral with respect to $s$ and then the integral with respect to $m$, one obtains the contribution to the effective action
\begin{equation}
\mathcal{L} \supset \frac{i}{16 \pi^{2}} F_{\mu \nu} \widetilde{F}_{\mu \nu} \sum_{n = - \infty}^{\infty} \arctan{\left(\frac{m}{\omega_{n} - \frac{a}{2\pi R}} \right)}  \,.
\label{piece2}
\end{equation}
By taking a derivative with respect to $a$, we arrive at expression in equation~\eqref{momentumint} and thus find another representation of $g(a)$ as an infinite series. 

The sum of equations~\eqref{piece1} and \eqref{piece2} provides us with an alternative worldline formulation resulting in an effective axion-Euler-Heisenberg Lagrangian providing us with the effective low energy axion Maxwell theory Lagrangian to all orders in the axion and photon.

\section{Kaluza-Klein calculation} \label{appendix1}
In this appendix we re-derive the axion potential $V(a)$ (equation~\eqref{5dpotential}) and the effective axion coupling $g(a)$ (equation~\eqref{5dga}) for a tower of fermionic states in the Kaluza-Klein decomposition basis. We take the fermion to satisfy anti-periodic boundary conditions such that $\omega_{n} = \frac{\pi}{2 \pi R} \left(2n+1\right)$ in order to align the minimum of the potential with $a = 0$. The axion potential for other boundary conditions can always be obtained by shifting the axion $a$.

\subsection{The axion potential}
The axion gains an effective potential from interactions with the fermion. We can calculate this effective potential by taking the axion $a$ to be constant, 
\begin{equation}
V(a) =  -2  \sum_{n = - \infty}^{\infty} \int \frac{d^{4}p}{(2\pi)^{4}} \ln{\left( \left(\omega_{n} -  \frac{a}{2 \pi R }\right)^{2} + p^{2} + m^{2}\right)}  \,.
\label{potential5dappendix}
\end{equation}

The potential in equation \eqref{potential5dappendix} is well-known \cite{Sundrum:2005jf} and the resulting potential for the axion is
\begin{equation}
V(a) = -2 \int \frac{d^{4}p}{(2\pi)^{4}} \ln{\left(1+ e^{-4\pi R E_{p}} + 2 e^{-2 \pi R E_{p}}\cos{a} \right)}  \,.
\label{KKpotential}
\end{equation}

\subsection{The effective axion-photon coupling}
We can calculate $g(a)$, the effective axion-photon coupling, by taking a derivative of the effective action (equation \eqref{effectiveaction}) with respect to the axion, which will contain a term of the form
\begin{equation}
 i\frac{g'(a)}{16 \pi^{2}} F \tilde{F} \subset (2\pi R) \frac{\delta S_{\mathrm{eff}}}{\delta a}  \,.
\label{gofa}
\end{equation}
We keep both a constant axion $a$ and constant zero KK mode of the photon $F_{\mu \nu}$.

Proceeding with the calculation, from equation \eqref{effectiveaction}, we have that
\begin{equation}
(2 \pi R)\frac{\delta S_{\mathrm{eff}}}{\delta a} \supset -\mathrm{Tr}\left(i\gamma^{5} \frac{1}{\slashed D + m}\right)  \,.
\label{greensfunction}
\end{equation}
For a constant axion we can expand the denominator as
\begin{equation}
(2 \pi R)\frac{\delta S_{\mathrm{eff}}}{\delta a} \supset -\mathrm{Tr}\left(i\gamma^{5} \frac{\slashed D - m}{D^{2} - \frac{1}{2} \sigma^{\mu \nu} F_{\mu \nu} - m^{2}}\right)  \,.
\end{equation}
We proceed to expand the denominator to order $F^{2}$ as
\begin{equation}
(2 \pi R)\frac{\delta S_{\mathrm{eff}}}{\delta a} \supset - \frac{1}{4}\mathrm{Tr}\left(i\gamma^{5} \left( \slashed D - m\right)\left(\frac{1}{D^{2} - m^{2}}\right)^{3} \sigma^{\mu \nu} F_{\mu \nu} \sigma^{\alpha \beta} F_{\alpha \beta} \right)  \,.
\end{equation}
Using the identity $\mathrm{Tr}\left(\gamma^{5} \sigma_{\mu \nu} \sigma_{\alpha \beta}\right)= -4 \epsilon_{\mu \nu \alpha \beta}$ and ignoring higher order $F$ contributions, we find that

\begin{equation}
(2 \pi R)\frac{\delta S_{\mathrm{eff}}}{\delta a} \supset  \frac{2im}{2\pi R}F \widetilde{F} \sum_{n = - \infty}^{\infty} \int \frac{d^{4}k}{(2\pi)^{4}} \frac{1}{\left( \left(\omega_{n} - \frac{a}{2 \pi R }\right)^{2} + k^{2} + m^{2} \right)^{3}}  \,.
\end{equation}
Performing the momentum integral yields
\begin{equation}
(2 \pi R)\frac{\delta S_{\mathrm{eff}}}{\delta a} \supset   \frac{im}{2\pi R}F \widetilde{F}\sum_{n = - \infty}^{\infty} \frac{1}{16 \pi^{2}} \frac{1}{\left(\omega_{n}-\frac{a}{2 \pi R}\right)^{2} + m^{2} }  \,.
\label{momentumint}
\end{equation}
The frequency sum can be done by a method of images as
\begin{equation}
\frac{1}{2 \pi R}\sum_{n = - \infty}^{\infty}\frac{1}{\left(\omega_{n}- \frac{a}{2 \pi R }\right)^{2} + m^{2} } = \frac{1}{2|m|}\frac{\sinh{\left(2\pi R |m|\right)}}{\cosh{\left(2 \pi R m \right)}+\cos{\left(a\right)}}  \,.
\end{equation}
Plugging this into equation \eqref{momentumint}, one obtains
\begin{equation}
(2 \pi R)\frac{\delta S_{\mathrm{eff}}}{\delta a} \supset i  \frac{\sinh{\left(2\pi R |m| \right)}}{\cosh{\left(2 \pi R m \right)}+\cos{\left(a\right)}}\frac{\mathrm{sign}(m)}{32 \pi^{2}} F \Tilde{F}  \,.
\end{equation}
By comparing this expression with equation \eqref{gofa}, we see that
\begin{equation}
g'(a) = \frac{1}{2} \frac{\sinh{\left(2\pi R m\right)}}{\cosh{\left( \pi R m \right)}+\cos{\left(a\right)}}  \,,
\end{equation}
which implies
\begin{equation}
g(a) =  \arctan{\left( \tanh{\left( \pi R m\right)} \tan{\left(\frac{a}{2}\right)}  \right)} + \pi \mathrm{sign}(m)\Theta(a-\pi)  \,.
\end{equation}
This function has the correct $\frac{1}{2}$-level Chern Simons.

\bibliography{ref}
\bibliographystyle{JHEP}

\end{document}